\documentclass[11pt,fleqn]{article}

\usepackage{amsmath} 
\usepackage{amsfonts}
\usepackage{graphicx}
\usepackage{dcolumn}
\usepackage{upgreek}
\usepackage{latexsym,slashed}
\usepackage{amssymb,amsfonts}
\usepackage{cancel}
\usepackage{bm} 
\usepackage{amsmath,mathrsfs} 

\newcommand{\pacs}[1]{\smallskip\noindent{\sl PACS numbers: \hspace{0.3cm}#1}\par\bigskip\rm}

\newcommand{\address}[1]{\begin{center}\large #1\end{center}}

\def\beq{\begin{eqnarray}}
\def\eeq{\end{eqnarray}}

\def\R{{\hbox{{\rm I}\kern-.2em\hbox{\rm R}}}} 
\def\H{{\hbox{{\rm I}\kern-.2em\hbox{\rm H}}}} 
\def\N{{\hbox{{\rm I}\kern-.2em\hbox{\rm N}}}} 
\def\C{{\hbox{{\rm I}\kern-.6em\hbox{\bf C}}}} 
\def\Z{{\hbox{{\rm Z}\kern-.4em\hbox{\rm Z}}}} 
 
\newcommand{\ack}[1]{\par\section*{Acknowledgement} #1}

\catcode`\@=11
\@addtoreset{equation}{section}

\begin{document}
\tolerance=5000

\title{Quantum detectors  in generic non flat FLRW space-times} 

\author{
Yevgeniya  Rabochaya$\,^{(a)(b)}$\footnote{rabochaya@science.unitn.it}
 and
Sergio~Zerbini$\,^{(a)(b)}$\footnote{zerbini@science.unitn.it}}
\date{}
\maketitle

\address{$^{(a)}$ Dipartimento di Fisica, Universit\`a di Trento, \\
Via Sommarive 14, 38123 Povo, Italia}

\address{$^{(b)}$ Centro INFN-TIFPA, \\
Via Sommarive 14, 38123 Povo, Italia}

\medskip 
\medskip

\begin{abstract}

A  quantum field theoretical approach, in which a quantum probe is used to investigate the properties
generic non-flat FLRW space-times  is discussed. The probe is 
identified with a conformally coupled massless scalar  field defined on a space-time with horizon and the procedure to investigate the local properties is realized by the use of Unruh-DeWitt detector and by the evaluation of the regularized quantum fluctuations. In the case of de Sitter space, 
the coordinate independence  of our results is checked, and the Gibbons-Hawking temperature is recovered. A possible generalization to the electromagnetic probe is also briefly indicated.  
\end{abstract}


\section{Introduction}

 Hawking discovery of black hole radiation \cite{Haw} is considered one of the most important predictions of quantum field theory in 
curved space-time. The predicted effect is quite robust, see \cite{dewitt,BD,wald,fulling,igor,U,B}. 

Parikh and Wilczek \cite{PW}(see also \cite{visser,vanzo}) introduced a further approach, the so-called 
tunnelling method, for investigating Hawking radiation. Later, the  Hamilton-Jacobi tunnelling method \cite{Angh,tanu,mann,obr} was introduced. This method is covariant and can be generalized  to the dynamical case \cite{bob07,bob08,sean09,bob09,bob10,Vanzo12,bin}.

The aim of this paper is to continue the investigation of the local properties of  a generic FLRW space-time by making use of suitable quantum probe, along the line presented in reference \cite{noi11}. We recall that FLRW space-times may be regarded  spherically symmetric (dynamical) space-times, and a covariant formalism introduced by  Hayward is at disposal \cite{sean09,bob09,kodama}. This approach will permit the introduction  of Kodama vector and related observers. For the sake of completeness, first we review this  general formalism. 

To begin with, let us recall that any spherically 
symmetric metric can locally be expressed in the form
\beq
\label{metric}
ds^2 =\gamma_{ij}(x^i)dx^idx^j+ R^2(x^i) d\Omega^2\,,\qquad i,j \in \{0,1\}\;,
\eeq
where the two-dimensional metric
\beq d\gamma^2=\gamma_{ij}(x^i)dx^idx^j
\label{nm}
\eeq
is referred to
as the normal metric, $\{x^i\}$ are associated coordinates and
$R(x^i)$ is the 
areal radius, considered as a scalar field in the two-dimensional
normal space. 
Let us introduce the  scalar quantity  
\beq
\chi(x)=\gamma^{ij}(x)\partial_i R(x)\partial_j R(x)\,. \label{sh} 
\eeq 
The dynamical trapping horizon, if it exists, is located in
correspondence of 
\beq 
\chi(x)\Big\vert_H = 0\,, \label{ho} 
\eeq
provided that $\partial_i\chi\vert_H \neq 0$.
Hayward surface gravity associated with this dynamical horizon reads 
\beq
\kappa_H=\frac{1}{2}\Box_{\gamma} R \Big\vert_H\,. \label{H} 
\eeq 
In any  spherical symmetric space-time there exists the Kodama vector field $\mathcal K$, defined by
\beq 
\mathcal K^i(x)=\frac{1}{ \sqrt{-\gamma}}\,\varepsilon^{ij}\partial_j R\,,
\qquad \mathcal K^\theta=0=\mathcal K^\varphi \label{ko} \;. 
\eeq 
Kodama observers are characterized by the condition $R=R_0$.

Coming back to the Hamilton-Jacobi tunneling method,  we recall the semi classical emission rate reads 
\beq
\Gamma \propto |\mbox{Amplitude}|^2 \propto 
e^{-2\frac{\Im I}{\hbar}} \,. 
\eeq
with $\Im$ standing for the imaginary part, the appearance of the imaginary part due to the presence of the horizon. 
The leading term in the WKB approximation of the tunnelling probability reads
\beq
\Gamma \propto e^{- \frac{2\pi }{\kappa_H} \omega}\,, 
\eeq
 in which an energy $\omega$ of the particle, and the Hayward surface gravity evaluated at horizon  appear.
Here we recall the operational interpretation \cite{sean09}. We note that
static observers in static BH space-times become in the dynamical case Kodama observers whose velocity 
\beq
v^i_K=\frac{K^i}{\sqrt{\chi}}\,, \quad  \gamma_{ij}v^i_Kv^j_K=-1\,.
\eeq
Kodama observers are such that $R=R_0$, namely they have constant areal radius, and  
the energy measured by these Kodama observers at fixed $R_0$ is
\beq
E=-v^i_K\partial_i I=-\frac{K^i \partial_i I}{\sqrt{\chi_0}}=\frac{\omega}{\sqrt{\chi_0}}\,.
\eeq
As a result, the tunneling rate  can be rewritten as
\beq
\Gamma  \simeq e^{-\frac{2 \pi}{ \kappa_H}\,\sqrt{\chi_0} E}\simeq e^{-\frac{E}{ T_0}}\,,
\eeq
and the local quantity $T_0$ at radial radius $R_0$ is  invariant, since  it contains the invariant factor 
$\sqrt{\chi}$, and the Hayward surface gravity 
\beq
T_0=\frac{T_H}{\sqrt{\chi_0}}\,, \quad T_H=\frac{k_H}{2\pi}\,.
\eeq
In the static case $\chi=g^{rr}=-g_{00}$ and recalling   Tolman's theorem: for a gravitational system at 
thermal equilibrium, $T\sqrt{-g_{00}}=\mbox{constant} $, it follows that  $T_H=\frac{\kappa_H}{2\pi}$ is the intrinsic
temperature of the BH, namely the Hawking temperature. 

As an important example that illustrates the gauge independence of the formalism is the de Sitter space-time. We shall consider three patches, or coordinate systems.
The first is the static patch, namely
\beq
ds^2=-dt_s^2(1-H_0^2r^2)+\frac{dr^2}{(1-H_0^2r^2)}+r^2dS^2\,.
\label{ds}
\eeq
Here $R=r$, the horizon is located at $r_H=\frac{1}{H_0}$, and the  surface gravity  $\kappa_H=H_0$.   

The second patch is the one which describes a exponentially expanding FLRW flat space-time,
\beq
ds^2=-dt^2+ e^{2H_0 t}\left(dr^2+r^2dS^2 \right)\,.
\label{dsf}
\eeq
Here $R=e^{H_0 t}r$, the dynamical horizon is $R_H=\frac{1}{H_0}$, there is no Killing vector, but the Hayward surface gravity is again  $\kappa_H=H_0$.
Finally, there exists the so called global dS patch, a non-flat FLRW space-time
\beq
ds^2=-dt^2+ \cosh^2( H_0 t)\left(\frac{dr^2}{(1-H_0^2r^2)}+r^2dS^2 \right)\,.
\label{dsc}
\eeq
Here $R= \cosh( H_0 t) r$, and a straightforward calculation  gives again $R_H=\frac{1}{H_0}$ and $\kappa_H=H_0$.

In the dynamical case, but for slow changes in the geometry,  the question is: could possibly  the quantity  
$T_H=\frac{\kappa_H}{2\pi}$ be interpreted as a dynamical Hawking temperature? In our opinion a complete answer is still missing, see also \cite{bin} for a recent discussion. With regard to this issue, it should be  important to have a quantum field theoretical confirmation of the tunnelling results (see, for example \cite{moretti}).

As a first step toward this aim, we recall that in the flat  FLRW case,  a conformally coupled massless scalar  has been used as a quantum probe in order 
to investigate such space-times (see \cite{obadia08,noi11,casadio} and references therein). The aim of this paper is to extend the investigation to the non flat 
FLRW space-times, since, at least in the case of De Sitter space-time, there exists the  important example (\ref{dsc}) of non flat FLRW space-time, namely the global  de Sitter patch. 

The paper is organized as follows. In Section 2 we give a brief survey of a scalar quantization in a generic 
FLRW space-time. In Section 3,  we present the main formula of Unruh-DeWitt response function detector. In Section 4, the computation of the 
renormalized  quantum fluctuation is presented and the de Sitter case is discussed in detail. In  Section 5, the conclusions are reported, and in the Appendix A, an elementary derivation of the Wightman function is given.

\section{Conformal quantum probe in FLRW space-times}

In the following, we review the quantization of a conformally coupled massless scalar field in a generic FLRW space-time.
First, it is convenient to  introduce the conformal time $\eta$ by means $d\eta=\frac{d t}{a}$. Thus, we have 
\begin{equation}
ds^2=a^2(\eta)(-d\eta^2+d \Sigma_3^2)\,,\;,
\label{cf} 
\end{equation}
where the metric of the spatial section may be written
\beq
d\Sigma_3^2= \frac{dr^2}{1-kh_0^2 r^2}+r^2dS^2\,, 
\eeq
with $k=0, \pm1$, and $h_0$ is a mass or inverse lenght scale, related to the Scalar Ricci curvature, which reads
\beq
\mathcal R=6\left( \frac{a''}{a^3}+k\frac{h_0^2}{a^2} \right)\,, \quad a'=\frac{da}{d \eta}\,.
\eeq
A useful and equivalent form for the spatial section is 
\beq
d\Sigma_3^2= d\xi^2+ h_k^2(\xi)dS^2\,, 
\eeq
where  $H_0 \xi= \sin^{-1} h_0 r$ and 
\beq
h(\xi)_1=\frac{\sin h_0 \xi}{h_0}\,, \quad h(\xi)_0= \xi \,, \quad h(\xi)_{-1}=\frac{\sinh h_0 \xi}{h_0}\,.
\eeq
In the case of a free massless scalar field which is conformally coupled to gravity,  
the related Wightman function $W(x,x')=<\phi(x) \phi(x')>$ can be computed in an exact way. 
For the sake of completeness, we re-derive this well known result. 

The quantum field $\phi$  has the usual expansion
\beq
\phi(x)=\sum_{\alpha}f_{\alpha}(x)a_{\alpha}+f^*_{\alpha}(x)a^+_{\alpha}
\eeq
where the modes functions $f_{\alpha}(x)$ satisfy the conformally invariant equation ($\mathcal R$ being the Ricci curvature)
\begin{equation}
\left(\Box -\frac{\mathcal R}{6}\right) f_{\alpha}(x)=0\;.
\label{f} 
\end{equation} 
Defining the vacuum by $a_{\alpha}|0>=0$, the Wightman function turns to be 
\begin{equation}
W(x,x')=\sum_{\alpha} f_{\alpha}(x)f^*_{\alpha}(x')\,,
\label{w} 
\end{equation}
and it satisfies, with $x'$ fixed,
\begin{equation}
\left(\Box -\frac{\mathcal R}{6}\right)W(x, x')=0\;.
\label{f1} 
\end{equation} 
An elementary derivation is presented in Appendix A. Here we give another derivation based on conformal invariance. 

First we recall  that if we make a 
conformal transformation
\beq
ds^2=\Omega(x)^2 ds_0^2\,, \quad \phi=\frac{1}{\Omega}\phi_0\,,
\eeq
the related Wightman $W(x.x')=<\phi(x)\phi(x')>$ transforms as
\beq
W(x,x')=\frac{1}{\Omega(x)\Omega(x')}W_0(x.x')\,.
\label{wc}
\eeq
We are interested in the non flat case. It is sufficient to consider only the $k=1$ case. The $k=-1$ may be obtained by the substitution $h_0 \rightarrow i h_0$.

Recall that  the metric on a non  flat FLRW positive spatial curvature space-time is
\beq
ds^2=a^2(\eta)\left( -d\eta^2+d\xi^2+\frac{1}{h_0^2}\sin^2 h_0\xi dS^2_2 \right)\,.
\eeq
As a result, the above metric  is conformally related to a static Einstein space $R \times S_3$, with metric
\beq
ds_E^2= -d\eta^2+d\xi^2+\frac{1}{h_0^2}\sin^2 h_0\xi dS^2_2 \,,
\eeq
On the other hand, it is well known that a static Einstein space-time is conformally related to Minkowski space-time, since
\beq
ds_E^2=4\cos^2 (h_0\frac{\eta+\xi}{2})4\cos^2 (h_0\frac{\eta-\xi}{2})\left( -dt^2+dr^2+r^2 dS^2_2 \right)
\label{e}
\eeq
with  the Minkowski coordinates  given by
\beq
t \pm r=\frac{1}{h_0}\tan (h_0\frac{\eta \pm \xi}{2})\,.
\eeq
Due to the homogeneity and isotropy of FLRW space-times,  we may take $W(x,x')=W(x',x)=W(\eta-\eta', r-r')$, same radial separation. Thus, since the Minkowski Wightman function is known, making use of equations (\ref{wc})  and (\ref{e}), the Einstein space Wightman function turns out to be 
\beq
W_E(x,x')=\frac{h_0^2}{8 \pi^2}\,\frac{1}{\cos(h_0(\eta-\eta'))-\cos(h_0(\xi-\xi'))}\,.
\eeq
As a consequence, again making use of (\ref{wc}),  the Wightman function related to FLRW spherical spatial section  is given by
\beq
W(x,x')=\frac{h_0^2}{8 \pi^2 a(\eta) a(\eta')}\,\frac{1}{\cos(h_0(\eta-\eta'))-\cos(h_0(\xi-\xi'))}\,.
\eeq
Finally, the Wightman function related to FLRW   hyperbolic spatial section  can be obtained by the replacement
$h_0\to ih_0$, and reads
\beq
W(x,x')=-\frac{h_0^2}{8 \pi^2 a(\eta) a(\eta')}\,\frac{1}{\cosh(h_0(\eta-\eta'))-\cosh(h_0(\xi-\xi'))}\,.
\eeq
As a check,  the  Wightman function related to FLRW flat spatial section can be obtained taking the limit $h_0 \rightarrow 0$. Again for same radial separation
\beq
W(x,x')=\frac{1}{4 \pi^2 a(\eta) a(\eta')}\,\frac{1}{-(\eta-\eta')^2+(r-r')^2}\,.
\eeq
These results are in agreement with the ones obtained in Appendix A.

\section{The Unruh-DeWitt detector in non flat FLRW space-times}

The Unruh-DeWitt detector approach is a well known and used technique for exploring quantum field theoretical aspects in 
curved space-time. For a recent review see \cite{crispino}. Here, we review the basic formula following 
Ref. \cite{louko,noi11}.

The transition probability per unit proper time of the detector depends on the response function per unit proper time which, for radial trajectories, at finite time $\tau$ may be written as  $\Delta \tau = \tau-\tau_0 >0$  
\beq 
\dot{ F}(E,\tau) =\frac{1}{2\pi^2}\mbox{Re}\int_{0}^{\Delta \tau } 
ds e^{-i E s} W(\tau, \tau-s)\,,
\eeq
where $\tau_0$ is the detector proper time at which we turn on the detector, and $E$ is the energy associated with the excited detector state (we are considering $E>0$). The flat case has been already considered in several places (see, \cite{noi11}), thus we shall consider the  $k=1$ FLRW case, namely
\beq 
\dot{ F}(E,\tau) =\frac{\mbox{Re}}{16\pi^2}\int_{0}^{\Delta \tau} 
\frac{ds}{a(\tau)a(\tau-s)} \frac{h_0^2 e^{-i E s}}{ \cos h_0(\Delta \chi(s))-\cos h_0 (\Delta \eta(s)-i0)}
\eeq 
where $\Delta \chi (s)=\chi(\tau)-\chi(\tau-s)$, and $\Delta \eta (s)=\eta(\tau)-\eta(\tau-s)$.
The $i0$-prescription is necessary in order to deal with the second order pole at $s=0$. However, 
we will show in the next Section  that the leading singularity in the coincidence limit, namely for small $s$, is  
\beq
 W(\tau, \tau-s) \simeq-\frac{1}{4\pi^2 s^2}+ O(s^0)\,.
\eeq
As a result, one may try to avoid the awkward limit $\epsilon \rightarrow 0^+$ by omitting the $\epsilon$-terms but subtracting the leading pole at $s=0$ (see \cite{louko} for details), and  introducing the quantity

\beq
\Sigma^2(\tau,s) \equiv a(\tau)a(\tau-s)\frac{2}{h^2_0}\left(\cos h_0(\Delta \chi(s))-\cos h_0 (\Delta \eta(s))\right) \,, 
\eeq
 one can present the detector transition probability per unit time in the form
\beq 
\hspace{-10pt}\dot{ F}(E,\tau) =\frac{1}{2 \pi^2}\int_{0}^{\infty}ds\, \cos( E s)
\left(\frac{1}{\Sigma^2(\tau,s)} + \frac{1}{s^2}\right) + J_\tau(E) \,,
\label{L6}
\eeq 
where the ''tail'' or finite time fluctuating term is given by
\beq
J_\tau(E) := -\frac{1}{2 \pi^2}\int_{\Delta \tau}^{\infty}ds\,\frac{\cos(E\,s)}{\Sigma^2(\tau,s)}\,.
\label{L00}
\eeq

In the important stationary cases (examples are the static black hole, the FLRW de Sitter space), one has  $\Sigma(\tau,s)^2=\Sigma^2(s)=\Sigma^2(-s)$,  and Eq.~\eqref{L6} simply becomes
\beq 
\dot{ F}(E,\tau) =\frac{1}{4\pi^2}\,\int_{-\infty}^{\infty}ds\, e^{-i E s}\left( \frac{1}{\Sigma^2(s)}+\frac{1}{s^2}\right) + J_\tau=\dot{ F}(E) +J_\tau(E) \,. 
\label{Lr}
\eeq 
The first term is independent on $\tau$, and  all the  time dependence is contained only in the fluctuating tail.

\subsection{The de Sitter space in FLRW patches}

In order to check the coordinate independence (gauge-invariance) of our result, it is instructive to investigate the de Sitter space-time. With regard to the other two FLRW patches, they are the flat spatial section, physically relevant for  inflation and the positive spatial curvature patch. The flat case has already been considered, see for example \cite{noi11}. 
For the spatial curved patch,  we shall make use of the formula derived in the previous subsection.   

 Recall that in a generic non- flat FLRW space-time, the Kodama observer is given by
\beq
R(t)=R_0=\frac{a(t)}{h_0} \sin h_0 \chi,
\eeq
with constant $R_0$. For a radial trajectory, the proper time in the non-flat FLRW is
\beq
d\tau^2=a^2(\eta)(d\eta^2-d\xi^2)\,.
\eeq
Thus, on the Kodama trajectory
\beq
d \tau^2=dt^2\left(1-\frac{R_0^2 H^2(t)}{a^2(t)-R_0^2H^2_0}\right)\,.
\eeq

In the case of de Sitter,  we put $h_0=H_0$ and the flat case is simple and one has an explicit expression for $\eta(\tau)$ \cite{noi11}. In the non-flat case, 
also for de Sitter, it is not easy to get an explicit expression of the conformal time as a function of $\tau$. For this reason we consider the $R_0=0$ case, comoving Kodama observer.
Thus $d\tau=dt$. Furthermore, since $a(t)=\cosh H_0 t=\cosh H_0 \tau$, one has
\beq
\eta(\tau)=\frac{2}{H_0} \arctan e^{H_0 \tau}\,.
\eeq
We have to compute
\beq\label{tetta}
\Sigma^2(\tau,s)= \frac{2}{H^2_0}\cosh (H_0 \tau) \cosh H_0 (\tau-s)\left( \cos( H_0 \Delta \eta)-1 \right) \,.
\label{i}
\eeq
Making use of well known trigonometric identities,  a direct calculation leads to following results
\beq
H_0 \Delta \eta(\tau, s)= -2 \arctan\left(\frac{\sinh \frac{H_0 s}{2}}{\cosh H_0(\tau +\frac{s}{2})}\right)\,, 
\eeq
\beq
\sin^2 H_0\Delta\eta(\tau,s)=\frac{\sinh^2\frac{H_0 s}{2}}{{\cosh^2H_0(\tau +\frac{s}{2})}+{\sinh^2\frac{H_0 s}{2}}},
\eeq
\beq
\Sigma^2(\tau,s)= \frac{2}{H^2_0}\cosh (H_0 \tau) \cosh H_0 (\tau-s) \sin^2 (H_0\Delta\eta(\tau,s))\,.
\eeq
As a consequence, the invariant distance  (\ref{i}) can be re-written in the form
\beq\label{tetta1}
\Sigma^2(\tau,s)=-\frac{2}{H_0^2}\sinh^2\left(\frac{H_0 \,s}{2}\right)\,.
\eeq
Since  $\Sigma^2(\tau, s)= \Sigma^2(s)=\Sigma^2(-s)$, we may use (\ref{Lr}) and obtain, for $E >0$ and making use of Residue Theorem
\beq 
\dot{ F}(E)=\frac{1}{2\pi} \,\frac{E}{e^{\frac{2\pi E}{H_0}}-1}\,,
\eeq
which shows that the Unruh-DeWitt detector in the non flat FLRW de Sitter space detects a quantum system in thermal equilibrium at a temperature 
$T_0=\frac{H_0}{2\pi}$, Gibbons-Hawking result is recovered \cite{GH}. This is an important check of the approach, since it
shows the coordinate independence of the result for the important case of de Sitter space. 

The response function for unit proper time, in the stationary cases we have considered, gives information about the 
equilibrium temperature via the Planckian distribution. We also may argue as follow.
 In the stationary case, in the limit $\tau \rightarrow \infty$, one has 
\beq 
\dot{ F}(E)=\frac{1}{4\pi^2}\,\int_{-\infty}^{\infty}ds\, e^{-i E s}\left( \frac{1}{\Sigma^2(s)}+\frac{1}{s^2}\right)= \frac{1}{2\pi}\; \frac{E}{\exp\left(\frac{ E}{T_0}\right)-1} \,. 
\label{x}
\eeq
 Note also that in this case one has 
\beq 
\frac{\dot{ F}(E)}{\dot{F}(-E)}=e^{-\frac{ E}{T_0}} \,,\quad \dot{ F}(-E)-\dot{ F}(E)=\frac{E}{2\pi}\,.
\eeq
Viceversa, if the above relations hold then $\dot{ F}(E)$ is the Plank distribution.
Thus we may  define the local equilibrium temperature by means of  
\beq
T_0=\frac{E}{\ln\dot{ F}(-E)-\ln \dot{ F}(E) }\,.
\eeq
or 
\beq
T_0=\frac{E}{ \ln{\left(1+\frac{E}{2\pi (\dot {F})}\right)}}\,
\eeq
which shows in which sense a Unruh-DeWitt detector is a quantum thermometer.

\subsection{d-dimensional generalization }
In the de Sitter case, it is possible to generalize the computation of the response function per unit time to the massive non conformally coupled d-dimensional scalar field \cite{proco}. 

In this case one may directly obtain
\beq
\dot{F}_{d,\nu}(E)=\frac{H_0^{d-3}e^{\frac{\pi E}{H_0}}}{8 \pi^{(d+1)/2} \Gamma(\frac{d-1}{2})} 
|\Gamma( \frac{d-1}{4}+\frac{\nu}{2}+ i \frac{E}{2H_0}) \Gamma( \frac{d-1}{4}-\frac{\nu}{2}+i \frac{E}{2H_0})  |^2\,,
\label{dds}
\eeq
where 
\beq
\nu=\sqrt{\frac{(d-1)^2}{4}-\frac{m^2}{H_0^2}-\xi d(d-1)  }\,,
\label{nu}
\eeq
with $m^2$ mass of the scalar field and $\xi$ the coupling constant with the Ricci curvature, being the conformal coupling $\xi_c=\frac{d-2}{4(d-1)}$.
As a consequence, one has 
\beq
\dot{F}_{d,\nu}(-E)= e^{\frac{E}{T_0}} \dot{F}_{d,\nu}(E)\,.
\eeq
Furthermore, the massless conformally coupled case in d dimension corresponds to $\nu=\frac{1}{2}$. In this case, for $d=4$ one recovers
\beq 
\dot{ F_{4,\frac{1}{2}}}(E)=\frac{1}{2\pi} \,\frac{E}{e^{\frac{2\pi E}{H_0}}-1}\,.
\eeq

Furthermore, making use of the identity 
\beq
\Gamma(z)|^2|\Gamma(1-z)|^2=\frac{2\pi^2 |z|^2}{\cosh( 2\pi \mbox{Im} z)-\cos(2\pi \mbox{Re}z  )}\,,
\eeq
one also has for $d=3$ and  $\nu=\frac{1}{2}$
\beq 
\dot{F}_{3,\frac{1}{2}}(E)=\frac{1}{16}\left(1+\frac{E^2}{H_0^2}\right)\frac{1}{e^{\frac{2\pi E}{H_0}}+1}\,,
\eeq
in which the well known phenomenon of the inversion of the statistic for odd dimensional spaces is manifest.

The other interesting case is the minimally coupled massless scalar field, for which $\nu=\frac{d-1}{2}$. In particular for $d=4$, one has 
\beq 
\dot{F}_{4,\frac{3}{2}}(E)=\frac{H_0^2}{8\pi^3}\left(1+\frac{E^2}{H_0^2}\right)\frac{1}{E\left(e^{\frac{2\pi E}{H_0}}-1 \right)}\,.
\eeq
In this physical important case, it mimics the graviton,  the well known infrared problem associated with it appears in the bad  behavior 
for small $E$.

\section{ Quantum fluctuations }

Another proposal to detect local temperature associated with stationary space-time admitting an event horizon has been
put forward by Buchholz and collaborators \cite{buchholz} (see also \cite{binosi,Eme}). The idea may be substantiated by the following argument.

Let us start with a free massless quantum scalar field $\phi(x)$ in thermal equilibrium at temperature $T$ 
in flat space-time. It is well known that finite temperature field theory effects of this kind may be investigated by  considering the scalar field defined in the Euclidean manifold $S_1 \times R^3$, where one has introduced the imaginary time $\tau=-it$, compactified in the circle $S_1$, with period $\beta=\frac{1}{T}$ (see for example \cite{byt96}). 

Let us  consider the local quantity $<\phi(x)^2>$, the quantum fluctuation. Formally this is a divergent quantity, since one is dealing with product of a valued operator distribution in the same point $x$, and regularization and renormalization are necessary. A simple and powerful way to deal with a regularized quantity is to make use of zeta-function regularization 
(see for example \cite{haw76,eli94,byt96}, and references therein). Within zeta-function regularization, one has
\beq
<\phi(x)^2>=\zeta(1|L_\beta)(x)\,, 
\label{M}
\eeq
where $ \zeta(z|L_\beta)(x)$ is the analytic continuation of the local zeta-function associated with the operator $L_\beta$ 
\beq
L_\beta=-\partial^2_\tau-\nabla^2\,,
\eeq
defined on $S_1 \times R^3$. 
The local zeta-function is defined with $\mbox{Re}\, z$ sufficiently large by means 
\beq
\zeta(z|L_\beta)(x)=\frac{1}{\Gamma(z)}\int_0^\infty dt t^{z-1}K_t(x,x)\,,
\label{M1}
\eeq
where the heat-kernel on the diagonal is given by 
\beq
K_t(x,x)=<x|e^{-tL_\beta}|x> =\frac{1}{\beta(4\pi t)^{3/2}}\sum_n e^{-\frac{4\pi^2}{\beta^2}n^2} \,.
\eeq
In (\ref{M}) the analytic continuation of the local  $ \zeta(z|L_\beta)(x)$ appears and it is assumed that this analytical continuation is regular at $z=1$, which, as we shall see,  is our case. If the analytic continuation has a simple pole in $z=1$, 
the prescription has to be modified (see \cite{iellici,binosi}). 

A standard computation, which makes use of the Jacobi-Poisson formula leads to
\beq
K_t(x,x)=\frac{1}{(4\pi t)^2}\sum_n e^{-\frac{n^2 \beta^2 }{4t}}\,. 
\eeq
Let us plug this expression in (\ref{M1}). The term $n=0$ leads to a formally divergent  integral $\int_0^\infty dt (t^{z-3 })$, but this is zero in the sense of Gelfand analytic continuation, and it can be neglected. Thus, a direct computation gives the analytic continuation of the local zeta-function
\beq
\zeta(z|L_\beta)(x)=\frac{\Gamma(2-z)}{8\pi^2 \Gamma(z)}\left(\frac{\beta^2}{4}\right)^{z-2}\zeta_R(4-2z)\,,
\eeq
where $\zeta_R(z)$ is the Riemann zeta-function. It is easy to see that the analytic continuation of the local zeta-function is regular at $z=1$, and from  (\ref{M}), recalling that $\zeta_R(2)=\frac{\pi^2}{6}$, one has
\beq
<\phi(x)^2>=\frac{1}{12 \beta^2}=\frac{T^2}{12}\,. 
\label{M2}
\eeq
Thus, the zeta-function renormalized  vacuum expectation value of the observable $\phi^2$, the fluctuation, gives the temperature of the quantum field in thermal equilibrium, namely one is dealing with  a quantum thermometer.  

Motivated by this argument, let us consider again a conformal coupled scalar field in a FLRW non flat space-time. 
We have seen that the off-diagonal Wigthman function is
\beq
W(x,x')=<\phi(x)\phi(x')>=\frac{1}{4\pi^2}\frac{1}{\Sigma^2(x,x')}\,,
\eeq
where 
\beq
\Sigma^2(\tau,\tau-s)= a(\tau)a(\tau-s)\frac{2}{H^2_0}\left(-\cos H_0(\Delta \xi(s))+\cos H_0 (\Delta \eta(s))\right) \,, 
\eeq
with $a(\tau)$ being the conformal factor. In the limit $s \rightarrow 0$, formally one has 
\beq
<\phi(x)^2>=W(\tau, \tau)\,,
\eeq
but $W(\tau,\tau)$ is ill defined, and one has to regularize and then renormalize this object. In our case, 
we may make use of the simple point splitting regularization \cite{BD}, namely we consider $W(\tau, \tau-s)$ and evaluate  the limit 
$ s \rightarrow 0 $. 

To implement this limit procedure, one has to make use of several identities. For radial time-like separation, the starting point is  
\beq
a^2(\tau)\left(\dot{\eta}^2- \dot{\chi}^2 \right)= 1\,.
\label{normaliz}
\eeq
Taking  first and second derivatives with respect the proper time, one has 
\beq
\frac{\dot{a}}{a}+a^2\left( \ddot{\eta}\,\dot{\eta}-\ddot{\xi}\,\dot{\xi}\right)=0\,,
\eeq
and
\beq
a^2(\dot{\eta}\dddot{\eta}-\dot{\xi}\dddot{\xi}=-a^2(\ddot{\eta}^2-\ddot{\xi}^2)-\frac{\ddot{a}}{a}+3(\frac{\dot{a}}{a})^2\,.
\eeq
Making use of these identities,  a long but straightforward calculation leads to
\beq
\Sigma^2=-s^2\left(1+\frac{B}{12}s^2+O(s^4) \right)\,.
\eeq
where
\beq
B=H^2+A^2+2\dot{H}\dot{t}+\frac{h_0^2}{a^2}\left(1-2\dot{t}^2 \right)\,.
\eeq
In this expression $A^2$ is the square of the 4-acceleration along the time-like trajectory, given by 
\beq
A^2=\frac{1}{\dot{t}^2-1}\left(\ddot{t}+H( \dot{t}^2-1) \right)\,.
\label{a}
\eeq
In the above expression,  the last term is the new one with respect the flat case. Thus, the point splitting  gives
\beq
W(\tau,\tau-s)=-\frac{1}{4\pi s^2}+\frac{B }{48 \pi^2}+O(s^2)\,.
\eeq
With regard to the renormalization, we simply subtract the first divergent term for $s \rightarrow 0$. The physical meaning of this 
subtraction has been discussed in detail in reference \cite{noi11}, and it amounts to subtract the contribution related to an inertial trajectory in Minkowski space-time. Thus, the renormalized quantum fluctuation reads 
\beq
<\phi^2>_R=\frac{1}{48 \pi^2}\left(H^2+A^2+2\dot{H}\dot{t}+\frac{h_0^2}{a^2}\left(1-2\dot{t}^2 \right)\right)\,.
\eeq
This result is the  generalization of the one obtained in a flat FLRW space-time in \cite{noi11} and within Unruh-de Witt detector in \cite{obadia08}. 

As a first important check, let us consider again the de Sitter space-time in the global patch. In this stationary case, the fluctuation acts as a 
quantum thermometer. Recall that  in this case, one has $h_0=H_0$,
\beq
 a(t)=\cosh H_0 t \,,\quad
H(t)=H_0\frac{\sin h H_0 t}{\cosh H_0 t}\,, \quad \dot{ H}=\dot{ t} \frac{H_0^2}{(\cosh H_0 t)^2}\,. 
\eeq
As a consequence
\beq
<\phi^2>_R=\frac{1}{48 \pi^2}\left(H_0^2+A^2\right)\,.
\eeq
This is in agreement with the result obtained for the de Sitter space-time in reference \cite{thirr}. In fact, the acceleration can be computed, since, for a Kodama observer $R=R_0$, one has
\beq
\dot{t}^2-1=\frac{R_0^2 H^2(t)}{1-R_0^2H_0^2}\,.
\eeq
Thus
\beq
\ddot{t}=\frac{R_0^2}{1-R_0^2H_0^2} H(t)\frac{d H}{d t}\,.
\eeq
Taking  equation (\ref{a}) into account, one gets
\beq
A^2=\frac{R_0^2 H^4_0}{1-R_0^2H_0^2}\,,
\eeq 
which coincides with the Kodama de Sitter acceleration evaluated in the flat patch \cite{noi11}. Furthermore we also have 
\beq
<\phi^2>_R=\frac{1}{48 \pi^2}\frac{H_0^2}{1-R_0^2H_0^2}  \,.
\eeq
For a comoving Kodama observer $R_0=0$, and one gets again  Gibbons-Hawking temperature associated with de Sitter space-time.

\section{Conclusion}

In this paper, with the aim to better understand the temperature-versus-surface gravity paradigm, the asymptotic results obtained by tunnelling semi classical method have been tested with quantum field theory techniques like the Unruh-DeWitt 
detector and the evaluation of the quantum fluctuation associated with a massless conformally coupled scalar field. More precisely the results obtained in 
reference \cite{noi11} have been extended to a generic non spatial flat FLRW space-time. The de Sitter space-time in the global non flat FLRW patch has been used as important example, and the Gibbons-hawking temperature for the de Sitter space has been re-derived with our general fluctuation formula as well as the Unruh-de Witt detector technique.

With regard to possible generalization, at least in the flat case, our approach may also be extended to the Maxwell field as soon as one makes use of the result obtained in reference \cite{re}. In fact there,  quantizing the Maxwell field  in a flat FLRW space-time and in the so called W gauge, a conformal lifting of the Lorenz  gauge in Minkowski space-time, the Maxwell Wightman 
function has been obtained in the form
\beq
W_{\mu \nu}(x,x')=-\frac{1}{2}\left( \frac{g_{\mu \nu}(x)a(x')}{a(x)}+\frac{g_{\mu \nu}(x')a(x)}{a(x')} \right)W(x,x')\,,
\eeq  
where $W(x,x')$ is the Whightman associated with massless conformally coupled scalar field in flat FLRW space-time. This strongly suggests that our conformally coupled scalar probe may mimic quite well the  quantum  Maxwell field. Of course, working with the Maxwell field, the gauge invariant has to be implemented and the relevant quantity is the Wightman function associated to (for example) the magnetic field. This is a very interesting issue with important cosmological applications, see for example the recent  discussion appeared in \cite{campa,du}, and we hope to consider this issue in a future work.

\ack{We thank G.~Cognola, M.~Rinaldi, L.~Vanzo for several  discussions.}

\section{Appendix A}

In this Appendix an elementary derivation of Wightman function for a massless conformally coupled scalar field is presented. 
Due to the homogeneity and isotropy of FLRW space-times,  we may take $W(x,x')=W(x',x)=W(\eta-\eta', r-r')$. 

It is convenient to introduce the auxiliary quantity
\beq
Y(x,x')=a(\eta)a(\eta')W(x,x')\,,
\eeq
and, from the equation of motion in FLRW space-time with  the conformal time one has
\beq
-\frac{d^2 Y}{d \eta^2}-kH^2_0Y+\Delta_h Y=0\,,
\eeq
where $\Delta_h$ is the Laplace operator associated with the metric $d\Sigma_3^2$.

We may take $x'=0$. Let us start with the flat case $k=0$. One has
\beq
-\frac{\partial^2 Y}{\partial \eta^2}+\frac{\partial^2 Y}{\partial r^2}+2\frac{\partial Y}{\partial r} =0\,.
\eeq
The solution is
\beq
Y=\frac{1}{-\eta^2+r^2}
 \eeq
As a result, making use of the homogeneity and isotropy, and  dealing  with the   distribution nature of $W$,
\begin{equation}
W(x,x')= \frac{1}{4\pi^2 a(\eta)a(\eta')}\,\frac{1}{(r- r')^2-(\eta-\eta'-i\epsilon)^2 }\,.
\label{wold} 
\end{equation}
Here we leave understood the limit as $\epsilon\to 0^{+}$.

We may rewrite it in covariant form, according to Takagi \cite{T} and Schlicht \cite{S}  
 We adapt Schlicht's proposal to our conformally flat case, namely
\begin{equation}
W(x,x')=\frac{1}{4\pi^2 a(\eta)a(\eta')}\,\frac{1}{[( x- x')-i\epsilon (\dot{x}+\dot{x}')]^2 }\,.
\label{w3} 
\end{equation}
where an over dot stands for derivative with respect to proper time (see also  \cite{noi11}). 

Coming back to the non-flat case, it is sufficient to consider the positive curvature case $h_0> 0$. The negative one may be obtained with 
the replacement $h_0 \rightarrow ih_0$. \\ For the sake of simplicity, here we may take $h_0=1$, then dimensional analysis will give the 
complete expression.
Let us start with 
\beq
ds^2=a^2(\eta)\left(-d\eta^2+d\chi^2+ \sin^2 \chi dS^2 \right)\,.
\eeq
The Ricci scalar reads
\beq
\mathcal R=6 \frac{a''+A}{a^3}\,, \quad a'=\frac{da}{d \eta}\,.
\eeq
and the equation for $Y$ is
\beq
-\frac{\partial^2 Y}{\partial \eta^2}-Y+\frac{\partial^2 Y}{\partial \chi^2}+2 \frac{\cos \chi}{\sin \chi}\frac{\partial Y}{\partial \chi} =0\,.
\eeq
The solution of this partial differential equation is
\beq
Y=\frac{1}{\cos \eta -\cos \chi }\,.
\eeq
Making use of dimensional analysis one arrives at
\begin{equation}
W(x,x')= \frac{1}{8\pi^2 a(\eta)a(\eta')}\,\frac{h^2_0}{\cos h_0(\chi-\chi')-\cos H_0 (\eta-\eta'-i\epsilon) }\,.
\label{woldi} 
\end{equation}
In the case of negative spatial curvature one has
\begin{equation}
W(x,x')= -\frac{1}{8\pi^2 a(\eta)a(\eta')}\,\frac{h_0^2}{\cosh h_0(\chi-\chi')-\cosh h_0 (\eta-\eta'-i\epsilon) }\,.
\label{woldi1} 
\end{equation}

\end{document}